\documentclass[preprint,twocolumn,10pt,a4paper,byrevtex,prb,showpacs]{revtex4}
\usepackage{amssymb}
\usepackage{natbib}
\usepackage{graphicx}
\usepackage{amsmath}

\providecommand{\U}[1]{\protect\rule{.1in}{.1in}}

\begin{document}

\title{Band edge noise spectroscopy of a magnetic tunnel junction}
\author{Farkhad G. Aliev, Juan Pedro Cascales}
\affiliation{Dpto. Fisica de la Materia Condensada C-III, Instituto Nicolas Cabrera (INC) and Condensed Matter
Physics Institute (IFIMAC), Universidad Autonoma de Madrid, Madrid 28049, Spain}
\author{Ali Hallal and Mairbek Chshiev}
\affiliation{Univ. Grenoble Alpes, CNRS, INAC-SPINTEC, CEA, F-38000 Grenoble, France}
\author{Stephane Andrieu}
\affiliation{Institut Jean Lamour, Nancy Universit\`{e}, 54506 Vandoeuvre-les-Nancy
Cedex, France}
\date{\today }

\begin{abstract}
We propose a conceptually new way to gather information on the electron
bands of buried metal(semiconductor)/insulator interfaces. The bias
dependence of low frequency noise in Fe$_{1-x}$V$_{x}$/MgO/Fe (0 $<$ x $<$
0.25) tunnel junctions show clear anomalies at specific applied voltages,
reflecting electron tunneling to the band edges of the magnetic electrodes.
The change in magnitude of these noise anomalies with the magnetic state
allows evaluating the degree of spin mixing between the spin polarized bands
at the ferromagnet/insulator interface. Our results are in qualitative
agreement with numerical calculations.

\end{abstract}

\maketitle

Buried metal (semiconductor)/insulator interfaces are found at the heart of
electronics \cite{Kroemer2001}. The current in tunneling devices is
determined by the bias, barrier and density of states of the electrodes \cite%
{Tiusan2007,Yuasa07}. Electron states not allowed in bulk could become
permitted at the surface leading to topological \cite{Hasan2010,Schlenk2013}
or interface resonant states \cite{Belashchenko2005}. For metallic
structures the scarce knowledge on the interface bands is mainly obtained by
indirect methods such as ballistic electron emission spectroscopy \cite%
{Bell1988} or high-resolution X-ray spectroscopy \cite{Bonell2012}. The
possibility of a reliable and down-scalable in-situ methods to investigate
interface electron bands remains centrally important \cite{Berger2011}.

Tunneling magnetoresistance \cite{Moodera95,Miyazaki95,Julliere1975} is
extremely sensitive to the band structures of ferromagnet/insulator (FM/I)
interfaces \cite%
{deTeresa1999,Butler2001,Mathon2001,Bowen01,FVincent03,Parkin04,Yuasa04,Yuasa07,Stewart2010}%
. Despite recent attempts to understand the nature of the electron bands
which contribute to electron transport in spintronic devices \cite%
{Zermatten2008,Rungger2009, Harada2012}, the issue remains unsettled. The
main tool to characterize interfaces or barriers has been inelastic electron
tunneling spectroscopy (IETS) \cite{Jaklevic1966,Teixeira2012,Harada2012}
analyzing the derivative of the \emph{conductance as a function of bias}.
The resulting IETS signals depend on the tunneling density of states (DOS)
and inelastic scattering \cite{Tiusan2007,Yuasa07,Wortmann2005} which could
obscure the detection of the band edges in the presence of interface
disorder. The bias dependence of \emph{the conductance and its low frequency
fluctuations} could be an alternative way to study the interface or electron
confinement \cite{Nikolic1994,Xu2010} DOS.

A commonly accepted phenomenological approach relates the excess low
frequency noise (LFN), often inversely dependent on the frequency $f$, with
electrons scattering from defects characterized by a broad distribution of
relaxation times with energy \cite{Dutta1981}. If dominant defect states are
located close to the interfaces, they could create interface band edge tails
(supplemental Figure 1(a) or Fig.S.1(a)). Therefore, when the tunneling is
tuned to some specific band edge in the opposite electrode, the current
could acquire an extra LFN due to multiple relaxations originating from
defect states contributing to the formation of the band edge tails (Fig.S.1(b)).

In this Letter we investigate the bias dependence of conductance and LFN in
single barrier tunneling devices in order to determine in-situ the energies
of the band edges of the buried interfaces. We unambiguously demonstrate the
validity of the \emph{band edge noise spectroscopy (BENS)} concept by
studying seminal Fe/MgO/Fe MTJs with partial doping of the bottom electrode
(Fe) with Vanadium (V). Such substitution has been shown to reduce defect
states inside the MgO barrier due to improved interface matching between Fe$%
_{1-x}$V$_{x}$ and MgO in Fe$_{1-x}$V$_{x}$MgO/Fe MTJs. \cite%
{Bonell2009,Herranz2010,Bonell2010}. Our numerical simulations confirm that
tunneling of band-tail electrons, influenced by spin orbit interactions, are
responsible for the observed LFN anomalies.

Our magnetic tunnel junctions were grown by Molecular Beam Epitaxy (MBE) on
MgO (100) substrates under ultra high vacuum (typically $10^{-10}$ mbar)
conditions. Fe-V alloys were grown at room temperature by co-evaporation,
the layer being afterwards annealed up to 900K. The barrier thickness was
controlled by RHEED intensity oscillations. The MTJs were patterned by UV
photolithography and Ar etching to dimensions ranging from 10 $\mu m$ to 50 $%
\mu m.$ More details can be found in \cite{Bonell2009}. The noise
measurements setup was described earlier \cite{Guerrero06,Guerrero07}. The
typical noise power spectra (\emph{S}$_{V}$) in the antiparallel (AP) or
parallel (P) states reveal the presence of 1/f noise in the frequency range
between 1 and 50 Hz as $S_{V}(f)\propto 1/f^{\beta }$ (with $0.8<\beta <1.5$%
, see Fig.S.1(b). The bias dependence of the LFN\ has been determined
through the Hooge factor ($\alpha $) from the phenomenological expression: 
\emph{S}$_{V}$\emph{(f)=}$\alpha \cdot $\emph{(I}$\cdot $\emph{R)}$^{2}$%
\emph{/(A}$\cdot $\emph{f)}, where $R$,$I$, $A$ and $f$ indicate resistance, 
current, area and frequency, respectively. \cite%
{Guerrero07}. Qualitatively similar results have been obtained by analyzing
integrated LFN (Fig.S.1(c)). Shot noise (SN) was obtained from the frequency
independent part of the LFN below 10K \cite{Guerrero06}.

We begin by analyzing the electron transport and SN behavior at T=4K. The
zero bias TMR as a function of V content shows a maximum (Fig.1(a)),
confirming a reduction of the interface mismatch reported previously at room
temperature \cite{Bonell2009,Herranz2010,Bonell2010}. The nearly Poissonian
character of the tunneling statistics with Fano factor $F=1\pm 0.05$
(Fig.1(a)) indicates nearly direct tunneling processes.

Figure 1(b) shows the bias dependence of the Hooge factor $\alpha (V)$ in
both P and AP states for a Fe/MgO/Fe MTJ used as reference. One observes an
excess LFN below 200 mV, where FeO\cite{Du2010} and Fe/MgO\cite{Tiusan2007}
interface defect states have been predicted to influence the conductance.
For the MTJ with a non-optimised Fe/MgO interface one observes a strong
suppression of LFN with bias with weak anomalies in the $\alpha (V)$ around
0.5V, indicated by arrows.

\begin{figure}[tbp]
\begin{center}
\includegraphics[width=7cm]
{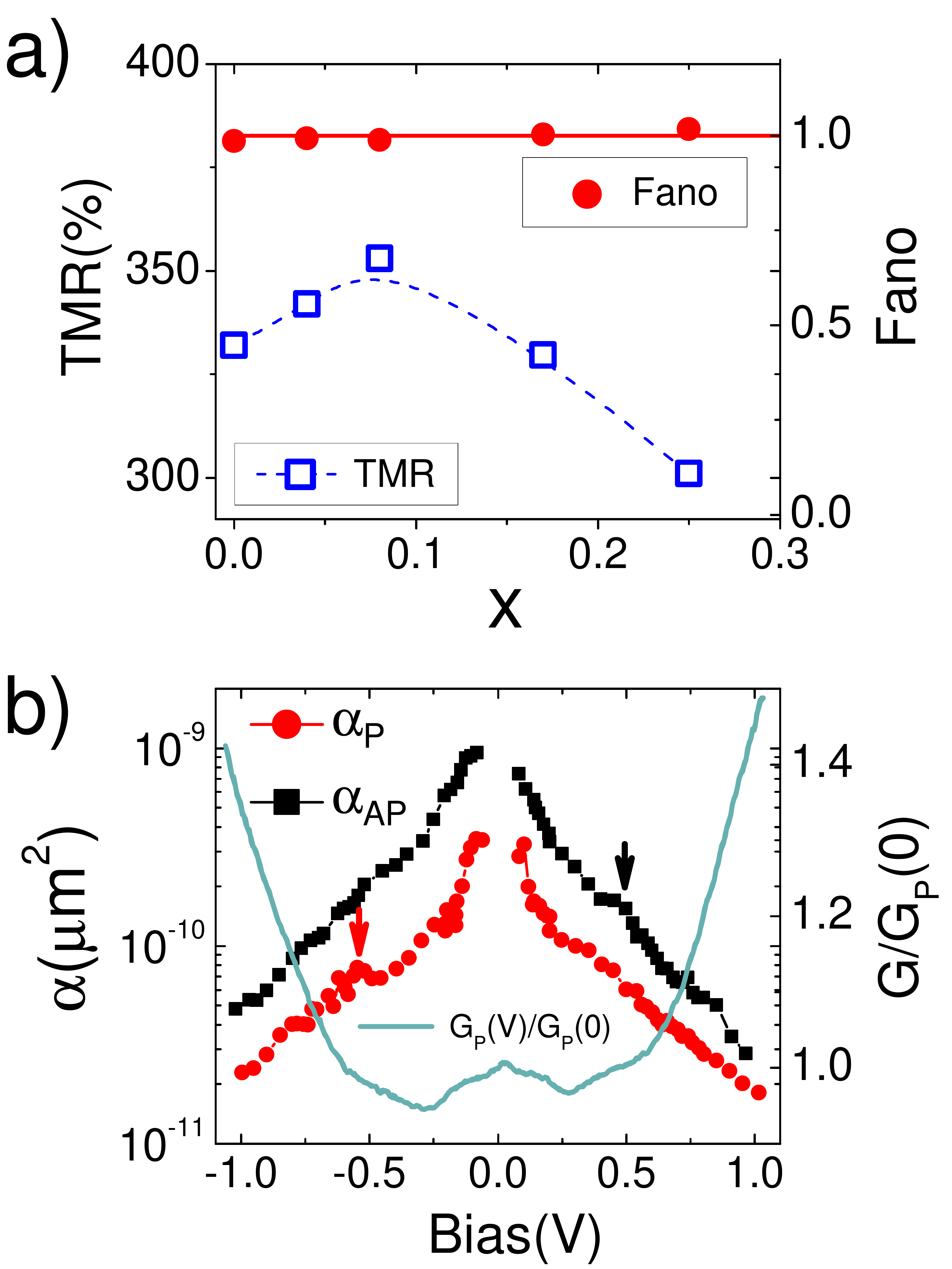} \label{Fig1}
\end{center}
\caption{(a) Dependence of the zero bias TMR and the Fano factor at $T=4K$
as a function of V content. (b) Bias dependence, at T=4K, of the dynamic
conductance in the P state, and the Hooge factor $\protect\alpha$ of both P
and AP states for Fe/MgO/Fe junctions. Arrows indicate weak peaks.}
\end{figure}

The doping of Fe with V improves the interface mismatch and decreases the
Fe/MgO interface defect states density \cite%
{Bonell2009,Herranz2010,Bonell2010}, which allows the implementation of the
BENS method. Figure 2(a) shows the $\alpha (V)$ and SN$(V)$ dependences for Fe$%
_{0.96}$V$_{0.04}$/MgO/Fe MTJs. The SN$(V)$ gives a Fano factor close to one,
proving direct tunneling in the bias range under study (Fig.2(a)). In
contrast to what is observed for the reference sample (Fig.1), the LFN shows
a clear enhancement (factor of 2) of conductance fluctuations around $\pm
0.6V$. Yet a stronger enhancement of the LFN close to 0.6V is observed in AP
configuration. The dynamic conductance in both states shows an upturn around
0.6V, but appears clearer in the P state (Fig.2, AP state not shown for
simplicity). Numerical calculations of the tunneling electron DOS indicate
that the upturn in conductance and the noise enhancement could be related
with the opening of a new transmission channel when the Fermi level of one
magnetic electrode crosses one of the band edges of the other magnetic
electrode, indicated by arrows in Fig.2(b).

\begin{figure}[tbp]
\begin{center}
\includegraphics[width=7cm]
{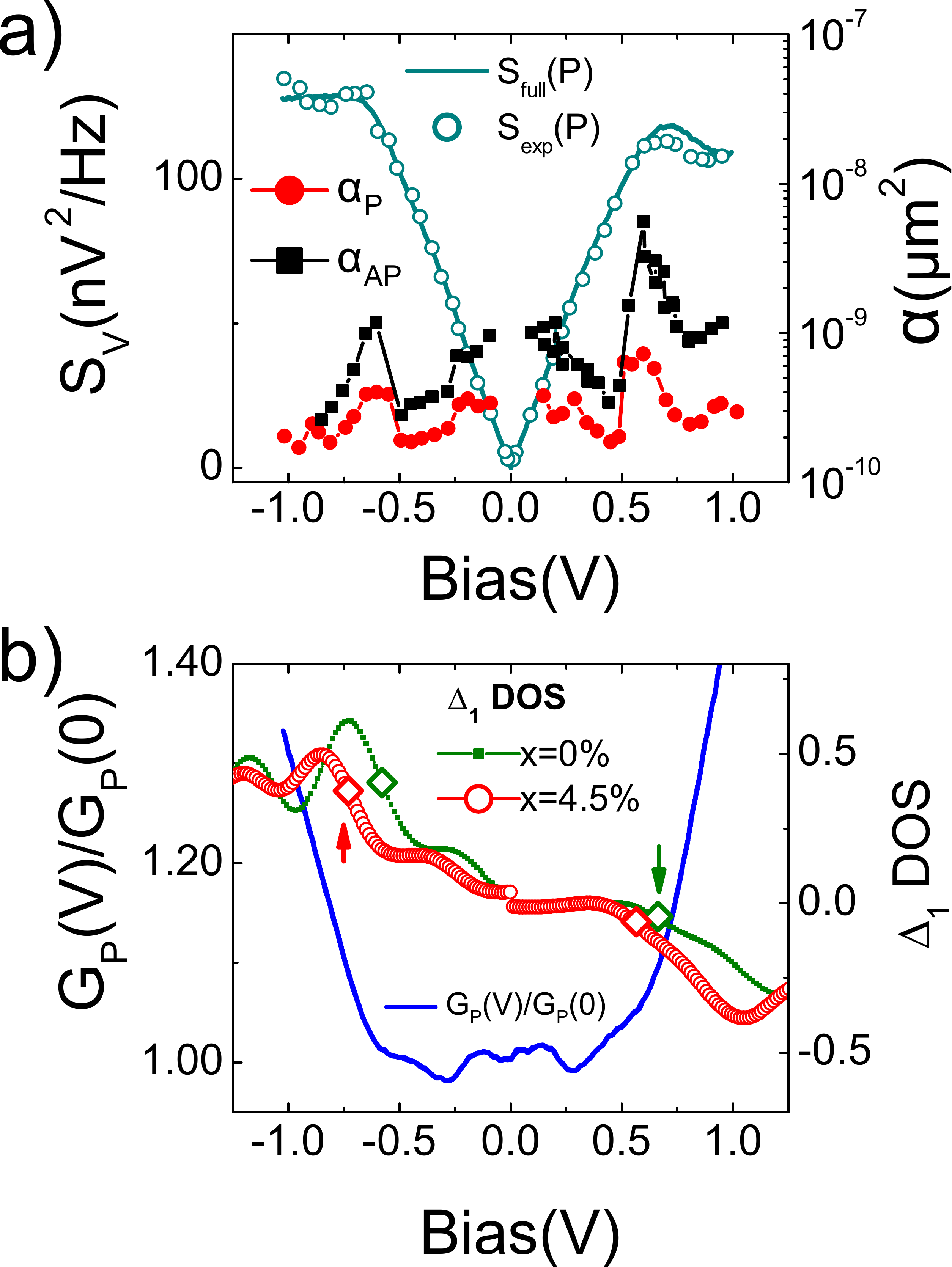} \label{Fig2}
\end{center}
\caption{(a) Bias dependence at $T=4K$ of the Hooge factor and SN for Fe$%
_{0.96}$ V$_{0.04}$ /MgO/Fe MTJ. (b) Dependence of the conductance with the
applied voltage at $T=4K$ combined with the calculated $\Delta_{1}$ DOS as a
function of energy with respect to $E_F$. Inflection points (open dots)
indicate $\Delta_1$ DOS band edges for 4\% Vanadium for $V<0$ and pure Fe
(x=0) for $V>0$.}
\end{figure}

Even clearer signs of the band edges in LFN are seen with an 8\% of V where the
lowest background LFN and the maximum TMR (Fig.1) are achieved. Figure 3
shows $\alpha (V)$ dependence in Fe$_{0.92}$V$_{0.08}$/MgO/Fe MTJs were the
optimum relation between two competing effects is reached: FM/I interface
relaxation on the one side and still not essential suppression of the
magnetization and the induced Fe-V structural disorder on the other side 
\cite{Bonell2009,Herranz2010,Bonell2010}. We estimate the TMR from our
simulations using the Julli\`{e}re model \cite{Julliere1975} (Fig.S.2) which
indicates the optimum values are reached for 9\% of V, i.e. rather close to what
is experimentally observed. We have found that the Fe$_{0.92}$V$_{0.08}$%
/MgO/Fe MTJs show clear anomalies in the Hooge factor for biases around 1V
and around 0.6V for the P state only, as shown in Fig. 3(a,b). Fig.3(d) demonstrates how the anomaly
in the P state around 0.6V gradually disappears with temperature, probably
due to thermal excitations.

\begin{figure}[tbp]
\begin{center}
\includegraphics[width=8cm]
{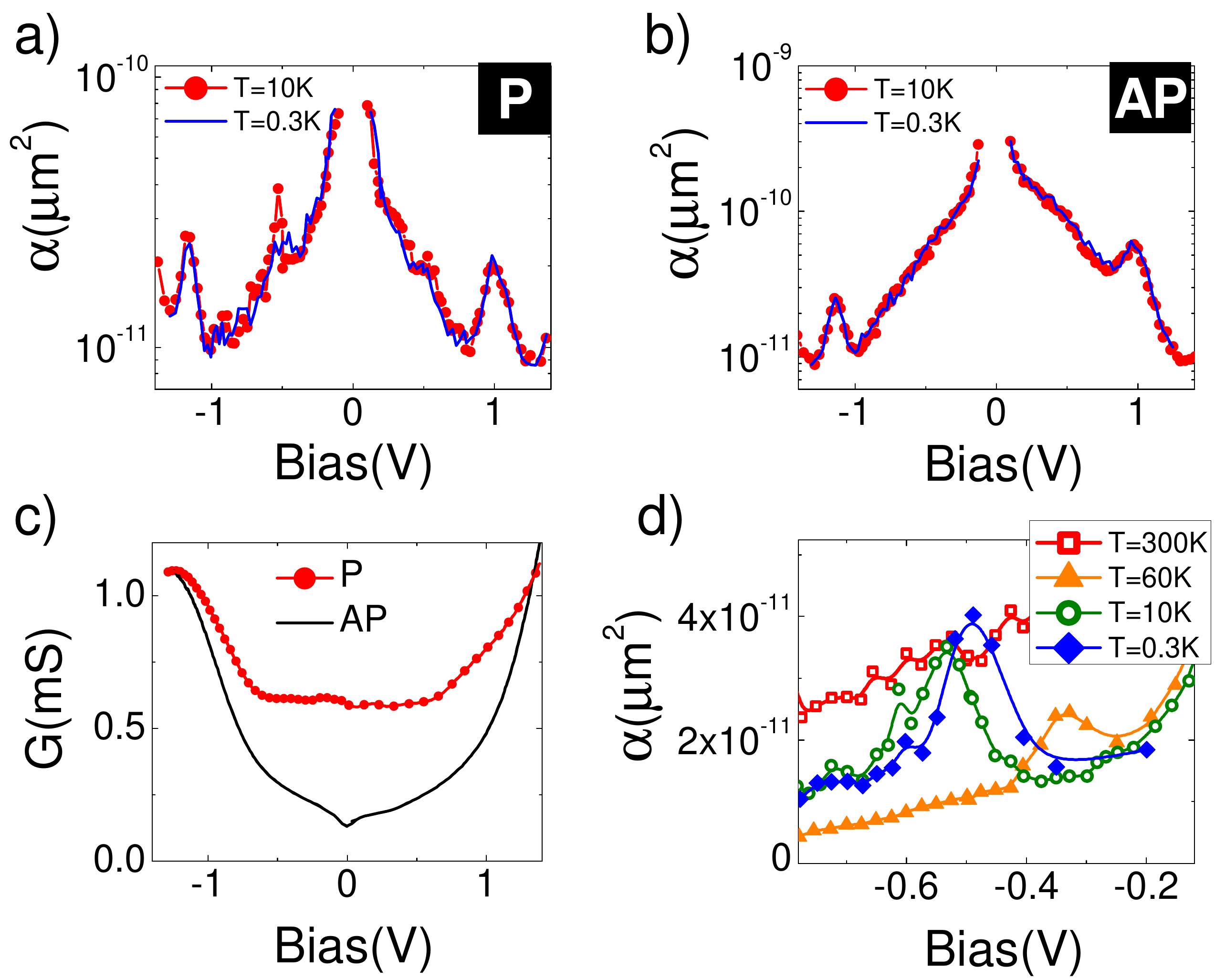} \label{Fig3}
\end{center}
\caption{Bias dependence at $T=4K$ of the Hooge coefficient for the (a) P
state and (b) AP state in Fe$_{0.92}$ V$_{0.08}$ /MgO/Fe MTJs. (c) Bias
dependence at $T=4K$ of the dynamic conductance for the P and AP state. (d)
Low frequency noise peaks gradually disappear with increasing temperature.}
\end{figure}

Qualitatively similar effects were seen for Fe$_{0.83}$V$_{0.17}$/MgO/Fe and
Fe$_{0.75}$V$_{0.25}$/MgO/Fe MTJs with the latter being the most
robust to electrical breakdown (standing up to 2.5V). In the high V content
range, the LFN is strongly influenced by random telegraph noise at positive
biases around 1V, reflecting a strongest asymmetry in interface defect
states previously visualized with scanning electron microscopy for Fe$_{0.8}$%
V$_{0.2}$/MgO/Fe MTJs \cite{Bonell2010}.

Fig.4(a) qualitatively explains the \emph{BENS} method. As long as tunneling
through the barrier is coherent, the main source for LFN are conductance
fluctuations due to atomic defects affecting $\Delta _{1}$\ and $\Delta _{5}$
interface states. Resulting localized states close to the band edges \cite%
{Fahy2006} could contribute, as reported for bulk semiconductors \cite%
{Jayaraman1989,Borovitskaya2001}, to the enhanced LFN. The key new feature
of the \emph{BENS} is the versatility in displacing the Fermi level ($E_{F}$
in Fig.4(a)) of the ejector electrode with respect to the different band
edges (or mobility edge, $E_{C}$ in Fig.4(a)) by simply varying the applied
bias. The right panel shows how the conductance and its derivatives are
expected to change when a new electron channel with a band edge opens at $%
E_{F}$. In order to clearly detect inelastic relaxation through IETS, some
well defined defect states should relax energy through coupling to a
well-defined set of phonon energies. We believe that the random interface
potential and the absence of well-defined defect states smear out the IETS
signals. Tunneling to the band tail weakly influences IETS (insert of Fig.S.1(c))
reflecting only the derivative of the DOS close to $E_{C}$. On the other
hand, much stronger changes in LFN vs. bias are seen due to a strong change
of excited defect relaxation times \cite{Borovitskaya2001} when tunneling
close to $E_{C}$, activating an excess of the low frequency conductance
fluctuations. Therefore, interface defect states dominate the LFN, and not
the derivative of the conductance (insert of Fig.S.1(c)).

The following arguments indicate that LFN\ mainly originates from
disorder/defects close to the FM/I interface: (i) direct tunneling (Fig.1);
(ii) the metallic nature of the electrodes, with resistance a few thousand
times below the barrier resistance, ensuring that electric signals and their
fluctuations mainly come from regions in the barrier and interfaces; (iii)
by analyzing LFN at higher biases we avoid direct resonant excitation of
localized FeO or O interface defect levels predicted below 200 mV \cite%
{Du2010}.

A simplified physical picture explaining the variation of LFN when tunneling to
three different energies $E_{1,2,3}$ around $E_{C}$ (Fig.4(a) and Fig.S.1) is as
follows. When electrons tunnel to energies $E_{1}>E_{C},$ their relaxation
time is fast due to the delocalized character of the band states near $E_{1}$
with a correspondingly small contribution to LFN. For tunneling to electron
states $E_{3}<E_{C}$ the LFN is also expected to be small due to low
probability of these tunneling events. However, when electrons tunnel to the
energies $E_{2}\lesssim E_{C}$, the tunneling current could be affected by
multiple trapping-type relaxations originating from shallow defect states
contributing to the formation of the band edge tails. We estimate that the
LFN peak width is roughly determined by the energy difference between the
mobility edge and the bottom of the band tail.

\begin{figure}[tbp]
\begin{center}
\includegraphics[width=8cm]
{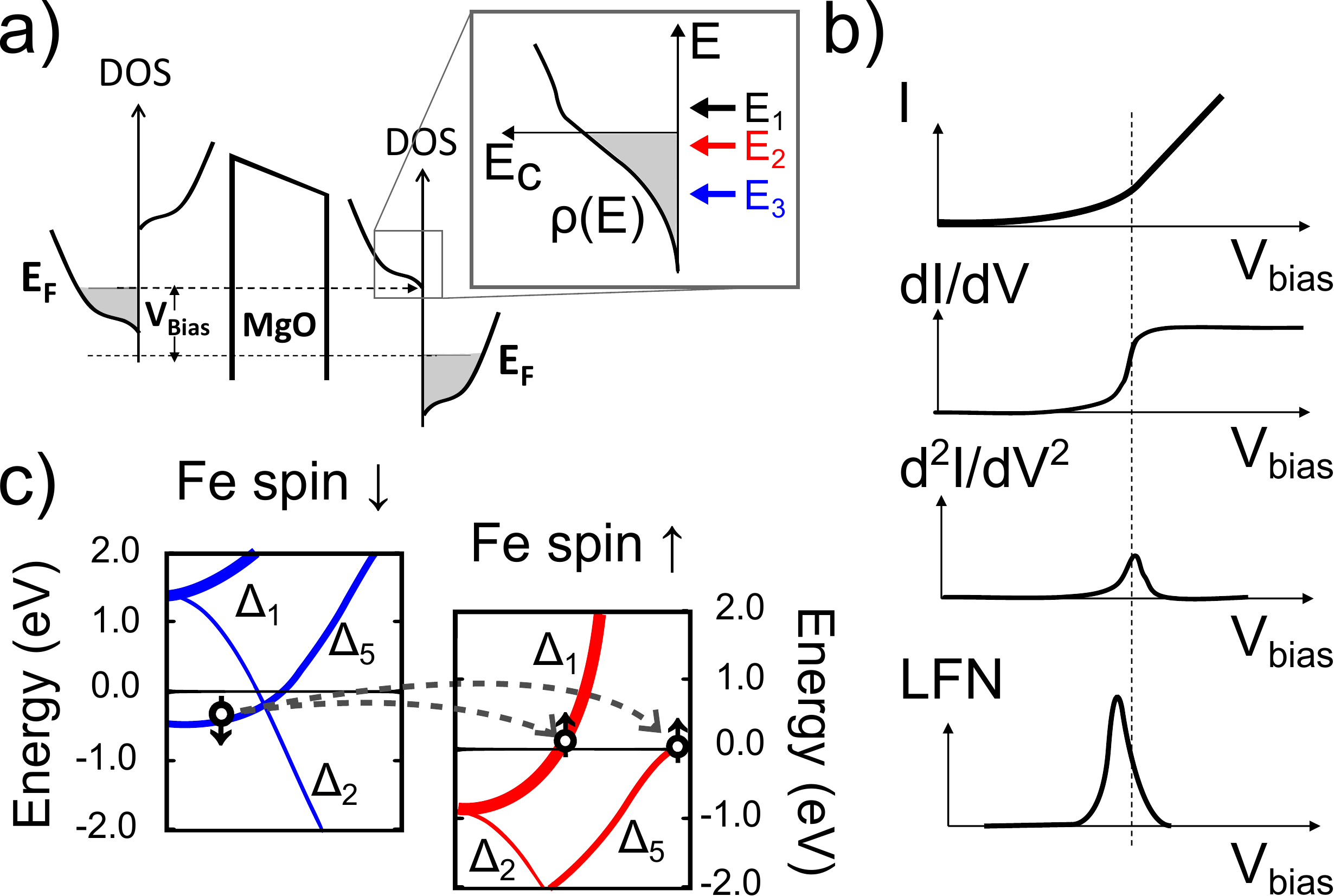} \label{Fig4}
\end{center}
\caption{(a) Sketch of the principle behind BENS, presented for the AP
state, where $E_C$ corresponds to the mobility edge. (b) The energies of
these defect states can be inferred from the IV curve of the sample, and its
first (dynamic conductance) and second (IETS) derivatives, but they are
detected in a much clearer way though low frequency measurements.(c) Sketch
of a band edge ($\Delta_{1},\Delta_{5}$) contribution to the tunneling at
$\sim -1.2V$.}
\end{figure}

In the MTJs under study, electron tunneling mainly occurs between polarized
bands with different Bloch state $\Delta _{1,5}$ symmetries spin filtered by
the MgO barrier \cite%
{Butler2001,Mathon2001,Bowen01,FVincent03,Parkin04,Yuasa04}. This allows a
rough estimation of the interband mixing at the interface by analyzing
variation of \emph{BENS} response with relative alignment of the electrodes.
Let us discuss qualitatively the reasons why \emph{BENS} could provide LFN
peaks both in the P and AP states (Figs. 2,3). For simplicity, we shall use
the majority and minority Fe electron bands tunneling in Fe/MgO/Fe junctions
(Fig.4(c)). When the MTJ is in the AP state, then in accordance with \emph{%
BENS} arguments $\Delta _{5\uparrow }\Rightarrow \Delta _{5\downarrow }$ and 
$\Delta _{1\uparrow }\Rightarrow \Delta _{1\downarrow }$ band edge tunneling
could provide a peak in LFN (AP) at different biases from 0.4 to 1.3 V if
conductance fluctuations originate from elastic scattering events.
Experimentally, however, we observe LFN peaks in the P state too (Fig.2(a)),
which we link with the presence of spin-orbit coupling induced $\Delta
_{1(\uparrow \downarrow )}\Longleftrightarrow \Delta _{5(\downarrow \uparrow
)}$ interband mixing at the Fe/MgO interface \cite{Lu2012}. Indeed, large
lateral momentum transfer and interband scattering could be dominant only
close to the interfaces \cite{Tsymbal2012}. Within such scenario, the
relation between amplitudes of the peaks LFN(P)/LFN(AP) provides an
evaluation of the degree of interband mixing between majority $\Delta
_{1\uparrow }$ band and the minority $\Delta _{5\downarrow }$ of roughly
0.2-0.3.

In order to examine quantitatively the applicability of our model we have
performed $ab-initio$ calculations of $\sqrt{2}\times \sqrt{2}$ unit cell of
Fe$_{1-x}$V$_{x}$ /MgO (x=0, 0.045, 0.091, 0.182) with a 5 monolayers (ML)
of MgO and 11 ML of Fe$_{1-x}$V$_{x}$. Our first-principles calculations are
based on density functional theory (DFT) as implemented in the Vienna $%
ab~initio$ simulation package (VASP)\cite{vasp} within the framework of the
projector augmented wave (PAW) potentials~\cite{paw} to describe
electron-ion interaction and generalized gradient approximation (GGA)\cite%
{gga} for exchange-correlation interactions. A 13$\times $13$\times $3
K-point mesh was used in our calculations. A plane wave energy cut-off equal
to 500 eV for all calculations was used and is found to be sufficient for
our system

Fig.5 compares the experimentally observed LFN anomalies in the P state
(open dots) with the band edge positions (closed dots) estimated from
inflection points in the DOS simulations for the majority and minority $%
\Delta _{1}$ and $\Delta _{5}$ states of Fe$_{1-x}$V$_{x}$/MgO (x=0, 0.045,
0.091, 0.182) structures (as indicated by arrows in the Fig.2(b)). We have
also indicated by horizontal dotted lines the estimated positions of the
band edges of the Fe/MgO structure.

\begin{figure}[tbp]
\begin{center}
\includegraphics[width=8.5cm]{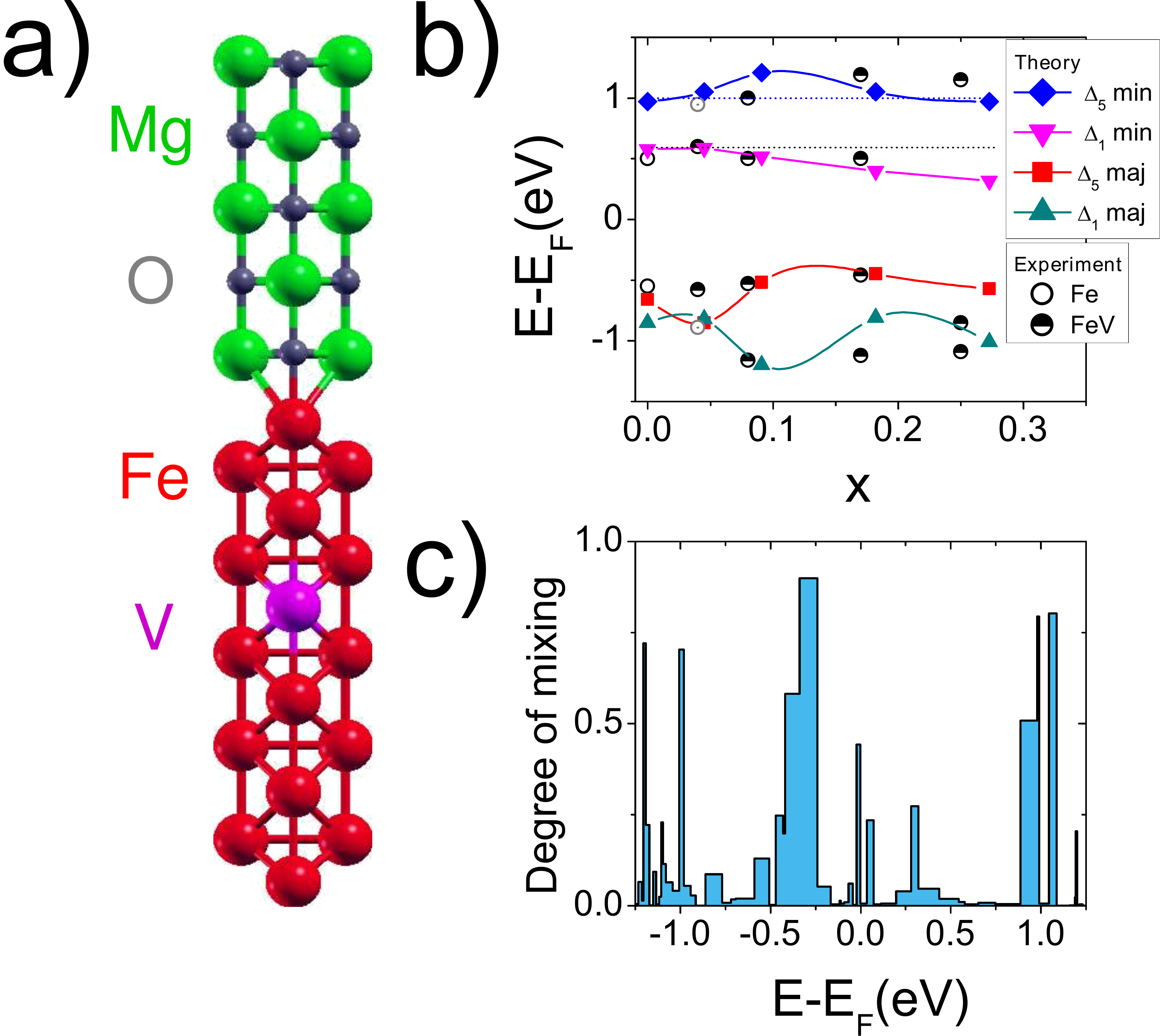} \label{Fig5}
\end{center}
\caption{(a) Schematic of the calculated crystalline structure for a $%
\protect\sqrt{2}\times\protect\sqrt{2}$ unit cell of (Fe$_{1-x}$V$_{x}$)$%
_{11}$/MgO$_{5}$ . (b) Calculated changes in the energies of the band edges
in Fe$_{1-x} $V$_{x}$ compared to the experimental data of low frequency
noise anomalies for the P state. Fully open experimental points indicate a
weak peak (increase of noise in less than 10\%). (c) Calculated degree of
mixing between $\Delta_1$ and $\Delta_5$ interface Bloch state character in
(Fe$_{1-x}$V$_x$)$_11$/MgO$_5$ for x=0.091.}
\end{figure}

A reasonable agreement between simulations and the experiment is observed,
especially for the Vanadium content between ($0.04<x<0.17)$ with reduced
lattice mismatch, the lowest background LFN and the highest TMR.

A few factors could contribute to some difference between experimental
results and calculations. First of all, calculations do not consider the
presence of dislocation induced mismatch as well as the structural disorder
difference between bottom and top interfaces\cite{Bonell2010}. On the
experimental side, measurements on MTJs with the least Vanadium were done
below 1V due to their vulnerability, making them difficult to compare with
the calculation results above 1V.

Finally, in order to better understand the influence of spin mixing at the
interface, we have also analyzed the Bloch state character of the
interfacial Fe atom in the presence of SOI as a function of the energy
difference to $E_{F}$. Fig.5(c) presents this analysis for $\Delta _{1}$ and 
$\Delta _{5}$ interface states in Fe$_{0.909}$V$_{0.091}$/MgO structure,
mainly participating in the electron tunneling through MgO. When the degree
of mixing at certain energy is equal to zero, it means that there is no
mixing between different $\Delta $ channels and there is only one $\Delta $
Bloch state character that dominates the tunneling at this energy tunneling.
The channel mixing is more pronounced at biases around $-(0.4\div 0.5)V$ and not
above $\pm 1V$, i.e. close to the intervals where LFN anomalies of different
magnitude were observed in both magnetic states (Fig.5(c)). We believe that $%
\Delta _{5\uparrow }\Rightarrow \Delta _{5\downarrow }$ and $\Delta
_{1\uparrow }\Rightarrow \Delta _{5\downarrow }$ mixing could be due to
surface induced band crossings and explains the appearance of peaks in LFN
both in the P and AP states.

\emph{To summarize}, we have introduced the band edge noise spectroscopy
concept which permits an investigation of the electron band edges in a wide
class of tunneling devices. We demonstrated successfully \emph{BENS}
approach in epitaxial magnetic tunnel junctions. The dependence of the \emph{%
BENS} on the relative magnetic alignment of the electrodes allows us to
estimate the importance of interband hybridization and spin flips at the
FM/I interfaces. Given the crucial importance of buried interfaces in
solid-state devices, the clear need to understand their electronic
structure, and the limited options available, our work presents a
substantial advance in the field of characterizing buried interfaces.

The authors acknowledge  A. Gomez-Ibarlucea, D. Herranz and F. Bonell for their help 
with the experiments and sample growth. This work has been supported by the 
Spanish MINECO (MAT2012-32743) and Comunidad de Madrid (P2009/MAT-1726) grants.


\begin{thebibliography}{99}
\bibitem{Kroemer2001} H. Kroemer, Rev. Mod. Phys. \textbf{73}, 783 ( 2001).

\bibitem{Tiusan2007} C Tiusan, et al, J. Phys.: Condens. Matter \textbf{19}
165201 (2007).

\bibitem{Yuasa07} S. Yuasa, D. D. Djayaprawira, J. Phys. D: Appl. Phys. 
\textbf{40} R337, (2007).

\bibitem{Hasan2010} M. Z. Hasan, C. L. Kane, Rev. Mod. Phys., \textbf{92},
3045 (2010).

\bibitem{Schlenk2013} T. Schlenk, et al., Phys. Rev. Lett., \textbf{110},
126804 (2013).

\bibitem{Belashchenko2005} K. D. Belashchenko, J. Velev, and E. Y. Tsymbal,
Phys. Rev. \textbf{B72}, 140404R (2005).


\bibitem{Bell1988} L. D. Bell, W.J. Kaiser, Phys. Rev. Lett., \textbf{61},
2368 (1988).

\bibitem{Bonell2012} F. Bonell, et al., Phys. Rev. Lett., \textbf{108},
176602 (2012)

\bibitem{Berger2011} R. F. Berger, C. J. Fennie, and J. B. Neaton, Phys.
Rev. Lett., \textbf{107}, 146804 (2011).

\bibitem{Julliere1975} M. Julli\`{e}re, Phys. Lett. \textbf{54A}, 225 (1975).

\bibitem{Moodera95} J.S. Moodera, L.R. Kinder, T.M. Wong, R. Meservey,
Phys.Rev.Lett. \textbf{74}, 3273 (1995).


\bibitem{Miyazaki95} T. Miyazaki, N. Tezuka, J. Magn. Magn. Mat. \textbf{139}%
, L231 (1995).

\bibitem{deTeresa1999} J. M. De Teresa, et al., Science, \textbf{286}, 507
(1999).

\bibitem{Butler2001} W. H. Butler, X.-G. Zhang, T. C. Schulthess, and J. M.
MacLaren, Phys. Rev. \textbf{B63}, 054416 (2001).

\bibitem{Mathon2001} J. Mathon and A. Umerski, Phys. Rev. \textbf{B 63},
220403(R) (2001).

\bibitem{Bowen01} M. Bowen, et al., Appl.Phys.Lett. \textbf{79}, 1655 (2001).

\bibitem{FVincent03} J.Faure-Vincent, et al., Appl.Phys.Lett. \textbf{82}
4507 (2003).

\bibitem{Parkin04} S.S.P. Parkin, et al., Nature Mat. \textbf{3}, 862 (2004).

\bibitem{Yuasa04} S. Yuasa, et al, Nature Mat. \textbf{3}, 868 (2004).

\bibitem{Stewart2010} D.A.Stewart, Nano Lett. \textbf{10}, 263 (2010).

\bibitem{Zermatten2008} P.-J. Zermatten, et al., Phys. Rev. \textbf{B78},
033301 (2008).

\bibitem{Rungger2009} I. Rungger,1 O. Mryasov, S. Sanvito, Phys. Rev. 
\textbf{B79}, 094414 (2009).

\bibitem{Harada2012} T. Harada, et al., Phys. Rev. Lett. \textbf{109},
076602 (2012).

\bibitem{Jaklevic1966} R. C. Jaklevic and J. Lambe, Phys. Rev. Lett. \textbf{%
17}, 1139 (1966).

\bibitem{Teixeira2012} J. M. Teixeira, et al., Phys. Rev. Lett. \textbf{106}%
, 196601 (2011).


\bibitem{Wortmann2005} D. Wortmann, H. Ishida, and S. Bl\"{u}gel, Phys. Rev. 
\textbf{B72}, 235113 2005

\bibitem{Nikolic1994} K. Nikolic and A. MacKinnon, Phys. Rev. B50, 11008 (\
1994),

\bibitem{Xu2010} G. Xu, et al., Nano Lett. \textbf{10}, 4590, (2010).

\bibitem{Dutta1981} P. Dutta and P.M. Horn, Review of Modern Physics, 
\textbf{53}, 497 (\ 1981).

\bibitem{Bonell2009} F. Bonell, et al . IEEE Trans. Mag.\textbf{\ 45,} 3467
(2009).

\bibitem{Herranz2010} D. Herranz, et al., Appl. Phys. Lett., \textbf{96},
202501 (2010).

\bibitem{Bonell2010} F. Bonell, et al., Phys. Rev. \textbf{B82}, 092405
(2010).

\bibitem{Guerrero06} R. Guerrero, et al., Phys. Rev. Lett., \textbf{97},
266602 (2006).

\bibitem{Guerrero07} R. Guerrero, et al., Appl. Phys. Lett., \textbf{91},
132504 (2007).

\bibitem{Du2010} G. X. Du, et al., Phys. Rev. \textbf{B81}, 064438 (2010).

\bibitem{Fahy2006} S. Fahy, A. Lindsay, H. Ouerdane, E. P. O'Reilly. Phys.
Rev. \textbf{B74}, 035203 (2006).

\bibitem{Jayaraman1989} R. Jayaraman, C.G. Sadini, IEEE Transactions on
Electronic Devices, \textbf{36}, 1773 ( 1989).

\bibitem{Borovitskaya2001} E. Borovitskaya, M.S.Shur, Sol. St. Electronics, 
\textbf{45}, 1067 (2001).

\bibitem{Lu2012} Y. Lu, et al., Phys. Rev. \textbf{B86}, 184420 (2012).

\bibitem{Tsymbal2012} E.Y.Tsymbal and I. Zitic, Handbook on spin transport
and magnetism, Ed., CRS Press, Taylor and Francis Group, 2012, p. 246.

\bibitem{Ando2005} Y. Ando, et al., Appl. Phys. Lett., \textbf{87}, 142502
(2005).

\bibitem{vasp} G.~Kresse and J.~Hafner, Phys. Rev. B \textbf{47}, 558
(1993); \textbf{54}, 11169 (1996); Comput. Mater. Sci. \textbf{6}, 15 (1996).

\bibitem{paw} P.~E. Bl\"ochl, Phys. Rev. B \textbf{50}, 17953 (1994); G.
Kresse and D. Joubert, Phys. Rev. B \textbf{59} 1758 (1999).

\bibitem{gga} Y.~Wang and J.~P.~Perdew, Phys. Rev. B \textbf{44}, 13298
(1991).
\end{thebibliography}
\end{document}